\documentclass{desyproc}

\begin{document}
\title{New results on possible higher twist contributions in proton diffractive structure functions at low x}

\author{{\slshape Mariusz Sadzikowski}\\[1ex]
Institute of Physics, Jagiellonian University, Reymonta 4,  30-059 Cracow, Poland
}

\contribID{smith\_joe}


\acronym{EDS'09} 

\maketitle

\begin{abstract}
The data on the diffractive deep inelastic scattering (DDIS) at HERA exhibit a strong excess, up to about 100\%,
above the twist-two NLO DGLAP description at low $Q^2$ and at large energy.
I show, that complementing the DGLAP fit by twist 4 and 6 components of the saturation model leads to a good description of data at low $Q^2$
and conclude that the DDIS at HERA provides the first evidence of higher twist effects in DIS.
\end{abstract}

\section{Introduction}

The diffractive DIS (DDIS) is a semi-inclusive process $ep\rightarrow epX$, in which proton
scatters elastically. Such processes create an important part (up to about 10 per cent) of the HERA events.
The fundamental description of DDIS is based on the leading twist contribution in which the large scale
is set by the negative four-momentum transfer $Q^2$ from the electron to the proton, carried by the
virtual photon. The proton structure functions $F_2^D, F_L^D$ are expressed in terms of
diffractive parton distribution functions (Dpdfs) due to the Collins factorization theorem \cite{collins}.
The dependence of Dpdfs on the hard scale $Q^2$ is governed by the DGLAP evolution equations.
Although the leading twist description of DDIS is successful I would like to point out in this presentation,
that the DGLAP fits fail to describe the DDIS cross-section HERA data below $Q^2=5$ GeV$^2$,
the problem that can be attributed to the negligence of the higher twist contributions.

\section{Cross-section and the DGLAP description}

The $t$-integrated  DDIS cross-section for the process
$e(k)p(P)\rightarrow e(k^\prime)p(P^\prime)X(P_X)$ reads:
\begin{equation}
\frac{d\sigma}{d\beta dQ^2 d\xi} = \frac{2\pi\alpha^2_{\mathrm{em}}}{\beta Q^4}[1+(1-y)^2]\sigma_r^{D(3)}(\beta, Q^2,\xi)
\end{equation}
where the invariants $y=(kq)/(kP)$, $Q^2=-q^2$, $\xi = (Q^2+M_X^2)/(W^2+Q^2)$ and $t=(P^\prime - P)^2$. The
quantity $W^2=(P+q)^2$ is the invariant mass squared in photon-proton scattering, and $M_X^2$ is the invariant mass
of the hadronic state $X$. The reduced-cross-section can be expressed in terms of the diffractive structure functions
$\sigma_r^{D(3)}(\beta, Q^2,\xi)=F_L^{D(3)}+F_T^{D(3)}$, whereas the structure functions $T,L$ are related to
transversally and longitudinally polarized $\gamma^\ast$ - proton cross sections
$F_{L,T}^{D(3)} = (Q^4/4\pi^2\alpha_{em}\beta\xi)d\sigma^{\gamma^\ast p}_{L,T}/dM^2_X$.

In the analysis \cite{ZEUS} the ZEUS diffractive data were fitted within NLO DGLAP approximation. A satisfactory
good description was found only for $Q^2>Q^2_{min}=5$ GeV$^2$. However, fits rapidly deteriorate with
decreasing $\xi$ and $Q^2$ reaching 100 percent effect at the minimal $Q^2=2.5$ GeV$^2$ and $\xi\simeq 4\cdot 10^{-4}$.
In the recent paper \cite{MSS} we confirmed this result (see Fig. 1, left panel, solid curve) and found that the
relative deviation from the DGLAP predictions exhibit, power like, $1/Q^2 - 1/Q^4$ dependence, which is
characteristic for higher twist effects.

\begin{figure}[t]
\centerline{\includegraphics[width=0.45\columnwidth]{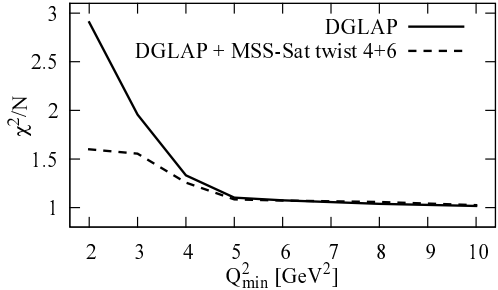}\hspace*{1.5cm}\includegraphics[width=0.45\textwidth]{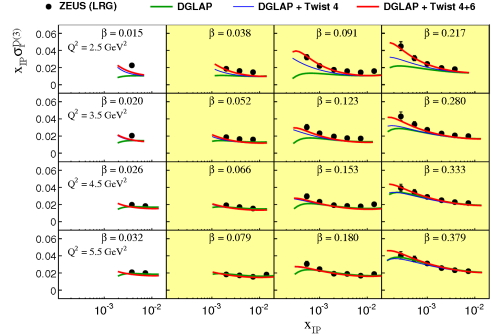}\vspace{0.5em}}
\caption{Left panel: the $ch^2$/d.o.f for NLO DGLAP and NLO DGLAP+HT fits to ZEUS LRG data \cite{ZEUS} with
 $Q^2<Q^2_{min}$. Right panel: The LRG ZEUS reduced cross-section data \cite{ZEUS} vs. DGLAP fit with included twist-4 and twist-4 and 6 corrections from
the MSS model \cite{MSS}.}\label{FigureLabel}
\end{figure}


\section{Higher twist contributions}

Our estimate of the HT contribution is based on the colour dipole model. In this approach the $\gamma^\ast p$ process is factorized
into an amplitude of photon fluctuation into the partonic debris and then scattering of these
states off the proton by the multiple gluon exchange. We take into account the contributions
from the fluctuation of the photon into a colour singlet quark-antiquark pair $q\bar{q}$ and
into $q\bar{q}$-gluon triple. The $t$-integrated $\gamma^\ast p$ cross section
$d\sigma_{L,T}^{\gamma^\ast p}/dM_X^2 = d\sigma_{L,T}^{q\bar{q}}/dM_X^2+d\sigma_{L,T}^{q\bar{q}g}/dM_X^2$.
Assuming an exponential $t$-dependence of diffractive cross-section, one finds for the $q\bar{q}$ component \cite{MSS}
\begin{equation}
\label{diff_cross_sec_qq}
\frac{d\sigma_{L,T}^{q\bar{q}}}{dM_x^2} = \frac{1}{16\pi b_D}\int\frac{d^2p}{(2\pi)^2}\int_0^1 dz \delta\left(\frac{p^2}{z\bar{z}}-M_x^2\right)
\sum_f\sum_{spin}\left| \int d^2r e^{i\vec{p}\cdot\vec{r}}\psi^f_{h\bar{h},\lambda}(Q,z,\vec{r})\sigma_d (r,\xi)\right|^2 .
\end{equation}
where $b_D$ is a diffractive slope, $z\bar{z}=z(1-z)$ and the first sum runs over the three light flavours.
The second sum of (\ref{diff_cross_sec_qq}) means summation over massless (anti)quark helicities $(\bar{h}) h$ in the case of longitudinal
photons whereas for transverse photons there is an additional average over initial photon polarizations $\lambda$. The squared photon
wave functions can be found, e.g. \cite{MKW} and we use the GBW parametrization for the dipole-proton cross section \cite{GBW}.
The contribution of the $q\bar{q}g$ component of $\gamma^\ast$ is calculated at $\beta =0$ and in the soft gluon approximation (the
longitudinal momentum carried by a gluon is much lower then carried by the $q\bar{q}$ pair). This approximation
is valid in the crucial region of $M_X^2\gg Q^2$ or $\beta \ll 1$, where the deviations from DGLAP are observed.
The correct $\beta$-dependence is then restored using a method described in \cite{marquet}. With these approximations one obtains:
\begin{eqnarray}
\label{cross_sec_qqg}
\frac{d\sigma^{q\bar{q}g}_{L,T}}{dM_x^2} &=& \frac{1}{16\pi b_D}\frac{N_c\alpha_s}{2\pi^2}\frac{\sigma_0^2}{M_x^2}\int d^2r_{01}
N^2_{q\bar{q}g} (r_{01},\xi)\sum_f\sum_{spin}\int_0^1 dz|\psi^f_{h\bar{h},\lambda}(Q,z,r_{01})|^2, \\\nonumber
N^2_{q\bar{q}g} (r_{01}) &=& \int d^2r_{02}\frac{r_{01}^2}{r_{02}^2 r_{12}^2}\left( N_{02}+N_{12}-N_{02}N_{12}-N_{01} \right)^2
\end{eqnarray}
where $N_{ij} = N(\vec{r}_j-\vec{r}_i)$, $\vec{r}_{01}, \vec{r}_{02},\vec{r}_{12}=\vec{r}_{02}-\vec{r}_{01}$ denote the relative positions of quark and antiquark $(01)$, quark and gluon $(02)$ in the transverse plain. The form of $N^2_{qqg}$ follows from the Good-Walker picture of the diffractive dissociation of the photon \cite{munier_shoshi}. The twist decomposition of (\ref{diff_cross_sec_qq}) is performed through the Taylor expansion in the inverse powers of $QR$ whereas
that of (\ref{cross_sec_qqg}) using Mellin transform technic \cite{MSS}.
From Fig. 1 (right panel), where we compare selected results with data, one can draw the the conclusions that a combination of the DGLAP fit,
twist-4 and twist-6 components of the model gives a good description of the data at low $Q^2$ much
better then a pure DGLAP fit extrapolation. Inclusion of
the higher twists terms improves the overall fit quality in the low $Q^2$ region (Fig. 1, left panel, dashed curve).
Nevertheless, it is important to stress that a truncation of the twist series (up to twist-6) is require
to have a good description of the data. We also compare our prediction with a recent H1 LRG measurement of Dpdfs \cite{H1_dpdfs}
which is given in Fig. 2 (left panel). Although we keep the parameters obtained in the fit of ZEUS data the
inclusion of HT corrections leads to better description of H1 data in compare to pure extrapolated DGLAP fits.

\begin{figure}[t]
\centerline{\includegraphics[width=0.4\columnwidth]{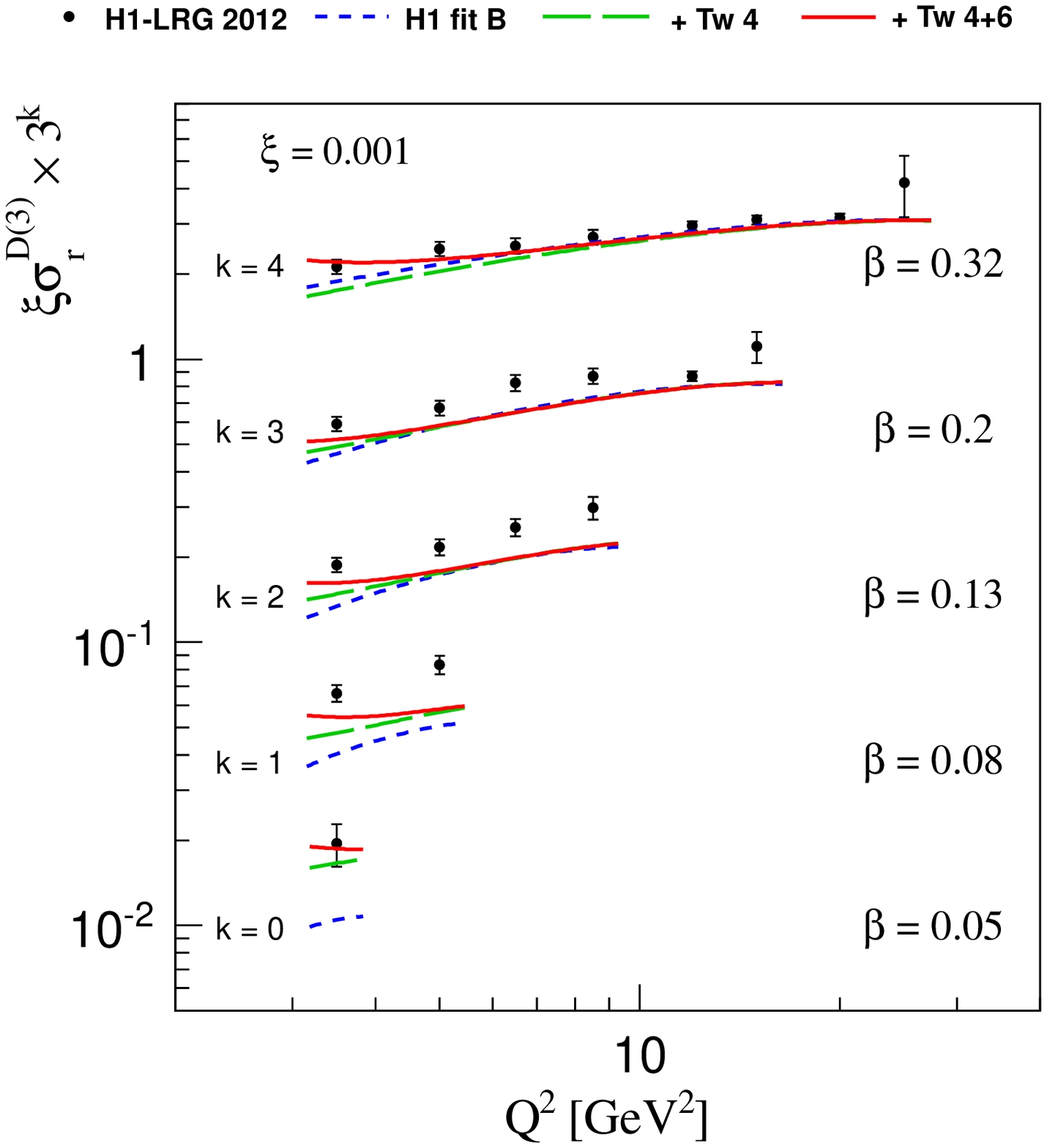}\hspace*{1.5cm}\includegraphics[width=0.45\textwidth]{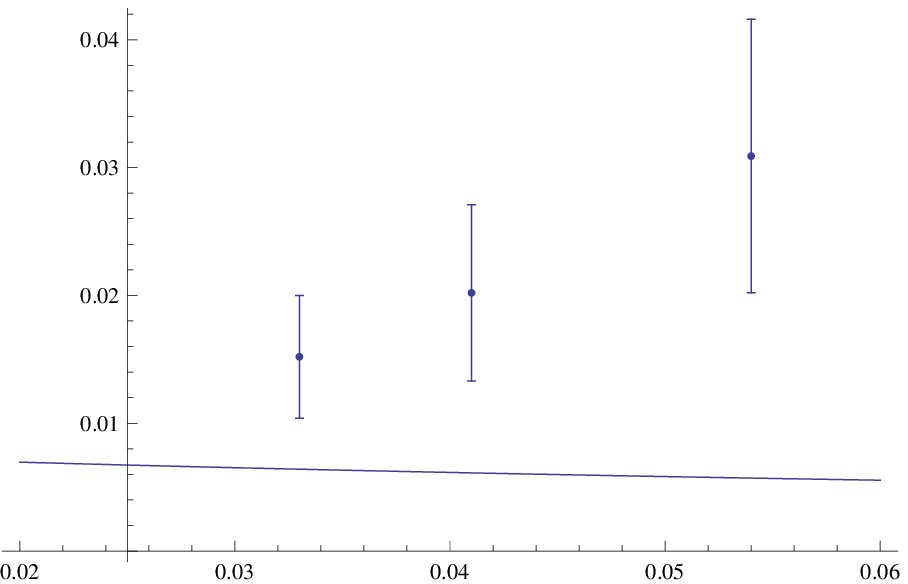}}
\caption{Left panel: The LRG H1 reduced cross-section data \cite{H1_dpdfs} at $\beta<0.4$. The dashed curve (the lowest) shows
the extrapolated H1 2006 DPDF fit B \cite{H1_dpdfs_fit} and the same fit with twist-4 contribution added (long-dashed line),
and with both twist-4 and twist-6 corrections (solid curve) from the MSS model \cite{MSS}. Right panel: Prediction of the MSS model (solid line) vs. H1 data points \cite{H1_FL} for the diffractive longitudinal structure function $F_L^D$ as a function of $\beta$ variable for bin $Q^2=4$ GeV$^2$ and $\xi = 0.003$.}\label{FigureLabel}
\end{figure}

It is a very interesting problem to compare our predictions with
the recent H1 measurement of the $F^D_L$ structure function \cite{H1_FL}. The preliminary result given in Fig. 2 (right
panel) clearly shows that the HT contributions do not lead to better description of data in compare to a pure
DGLAP fits (see \cite{H1_FL}) at low value of $Q^2$.

In conclusion one can state that the low $Q^2$ data on DDIS at HERA provides the best available ground for further
study of proton structure beyond the leading twist DGLAP desription.

\section{Acknowledgments}.

I would like to thank Organizers of EDS Blois 2013 conference for their hospitality.
This work is supported by the Polish National Science Center grant no. DEC-2011/01/B/ST2/03643.


\begin{footnotesize}

\end{footnotesize}
\end{document}